\documentclass{PoS}
\usepackage{amssymb,amsmath,graphicx}
\usepackage{slashed}

\title{Nucleon Charges, Form-factors and Neutron EDM}

\ShortTitle{Nucleon Charges, Form-factors and Neutron EDM}

\author{\speaker{Rajan Gupta} \footnote{LA-UR-15-28920} \\ 
	Theoretical Division, Los Alamos National Laboratory, Los Alamos, NM 87545, USA \\
	E-mail:\email{rajan@lanl.gov}}

\author{{Tanmoy Bhattacharya}\\
	Theoretical Division, Los Alamos National Laboratory, Los Alamos, NM 87545, USA \\
	E-mail:\email{tanmoy@lanl.gov}}

\author{{Vincenzo Cirigliano}\\
	Theoretical Division, Los Alamos National Laboratory, Los Alamos, NM 87545, USA \\
	E-mail:\email{cirigliano@lanl.gov}}

\author{{Huey-Wen Lin} \\
	Department of Physics, University of Washington, Seattle, WA 98195\\
	E-mail:\email{hwlin@phys.washington.edu}}

\author{{Boram Yoon} \\
	Theoretical Division, Los Alamos National Laboratory, Los Alamos, NM 87545, USA\\
	E-mail:\email{boram@lanl.gov}}

\abstract{We present an update of our analysis of statistical and
  systematic errors in the calculation of iso-vector scalar, axial and
  tensor charges of the nucleon. The calculations are done using
  $N_f=2+1+1$ flavor HISQ ensembles generated by the MILC
  Collaboration at three values of the lattice spacing
  ($a=0.12,\ 0.09,$ and $0.06$ fm) and three values of the quark mass
  ($M_\pi \approx 310,\ 220$ and $130$ MeV); and clover fermions for
  calculating the correlation functions, i.e., we use a clover-on-HISQ
  lattice formulation. The all-mode-averaging method allows us to
  increase the statistics by a factor of eight for the same
  computational cost leading to a better understanding of and control
  over excited state contamination.  Our current results, after
  extrapolation to the continuum limit and physical pion mass are
  $g_A^{u-d} = 1.21(3)$, $g_T^{u-d} = 1.005(59)$ and $g_S^{u-d} =
  0.95(12) $.  Further checks of control over all systematic errors,
  especially in $g_A^{u-d}$, are still being performed. Using results
  for the flavor-diagonal charges, $g_T^{u} $, $g_T^{d} $ and
  $g_T^{s}$, we analyze contributions of the quark electric dipole
  moment to the neutron EDM and the consequences for split SUSY
  model. }

\FullConference{The 33nd International Symposium on Lattice Field Theory\\
   14-18 July 2015\\
   Kobe International Conference Center, Kobe, Japan}

\begin{document}

\vskip -20pt
\section{Introduction}

The all-mode-averaging method allows us to obtain high
precision estimates for the matrix elements of flavor-diagonal and
isovector bilinear quark operators within nucleon states. These are
needed to probe many exciting areas of the Standard Model (SM) and its
extensions.  In Ref.~\cite{Bhattacharya:2011qm}, we showed that new
scalar and tensor interactions at the TeV scale could give rise to
corrections at the $10^{-3}$ level in precision measurements of the
helicity flip parts of the decay distribution of (ultra)cold neutrons
(UCN). In Ref.~\cite{Bhattacharya:2015wna}, we described the calculation of 
flavor-diagonal tensor charges $g_T^{u} $, $g_T^{d} $ and $g_T^{u+d} $
and analyze constrains on BSM theories using the quark EDMs and 
the current bound on the neutron EDM~\cite{Bhattacharya:2012nEDM}.

In these proceedings, we first describe the level of control achieved over
systematic errors, in particular the excited state contamination
(ESC), using $O(50,000)$ measurements with the all-mode-averaging
(AMA) method~\cite{Blum:2012uh}. We then summarize our results for the
iso-vector charges, the calculation of the flavor-diagonal tensor
charges $g_T^{u} $, $g_T^{d} $ and $g_T^{u+d} $ and an analysis the
contribution of quark EDM to the neutron EDM.  These calculations were
done using 9 ensembles of 2+1+1 flavor HISQ lattices generated by the
MILC collaboration~\cite{Bazavov:2012xda}. The matrix elements are
calculated using clover valence quarks on these HISQ ensembles.  A
summary of the parameters of the nine HISQ ensembles analyzed and the
number of measurements made in the fully high precision (HP) and the
AMA calculations is given in Table~\ref{tab:ensembles}.


\vskip -30pt
\section{All-mode-averaging Method}
\label{sec:AMA}

The all-mode-averaging (AMA) technique~\cite{Blum:2012uh} allowed us
to significantly increase the statistics very economically.  The basic
idea of the method is that one can construct correlation functions
using quark propagators (inverse of the Dirac matrix) calculated with
a low precision (LP) inversion criteria. The resulting bias can then be removed by
calculating the difference between correlated HP and LP estimates using much fewer 
source positions. The unbiased estimate for the two
($C_\text{LP}^\text{2pt}$) and three ($C_\text{HP}^\text{2pt}$) point
functions are given by
\begin{align}
 C^\text{imp} 
 = \frac{1}{N_\text{LP}} \sum_{i=1}^{N_\text{LP}}
    C_\text{LP}(\mathbf{x}_i^\text{LP}) 
  + \frac{1}{N_\text{HP}} \sum_{i=1}^{N_\text{HP}} \left[
    C_\text{HP}(\mathbf{x}_i^\text{HP})
    - C_\text{LP}(\mathbf{x}_i^\text{HP})
    \right] \,,
  \label{eq:2-3pt_AMA}
\end{align}
with $N_{\rm HP} \ll N_{\rm LP}$. Here  $\mathbf{x}_i^\text{LP}$ and $\mathbf{x}_i^\text{HP}$ are the
two kinds of source positions on each configuration from which the LP
and the HP correlators are calculated. Our AMA analysis done on five ensembles is described in
detail in~\cite{Bhattacharya:2015wna}. We use $64+4$ LP and 4 HP measurements on each
configuration. These 4 HP source calculations are the same as used in
the full high precision (HP) study presented
in~\cite{Bhattacharya:2015wna} and, therefore, needed no additional
calculations. In total, the new simulations generated $4 \times 16 +
4=68$ LP two- and three-point correlation functions per configuration.
The statistics used in the HP and the AMA analyses are given in Table~\ref{tab:ensembles}.

\begin{table}
\centering
\begin{tabular}{|ccccccc|c|}
\hline
Label    & $L^3\times T$   & $M_\pi$ MeV  & $(M_\pi L)$ & $N_\text{cfgs}$ & $N_\text{HP}$ & $N_\text{AMA}$ & $t_\text{sep}$ \\
\hline
a12m310  &$24^3\times 64$  & 305.3(4)     & $4.54$   &  1013   &  8104    & 64832 & 8, 9, 10, 11, 12  \\
a12m220S &$24^3\times 64$  & 218.1(4)     & $3.22$   &  1000   &  12000   &       & 8, 10 12          \\
a12m220  &$32^3\times 64$  & 216.9(2)     & $4.3$    &  958    &  7664    &       & 8, 10, 12         \\
a12m220L &$40^3\times 64$  & 217.0(2)     & $5.36$   &  1010   &  8080    & 68680 & 8, 10, 12, 14     \\
\hline
a09m310  &$32^3\times 96$  & 312.7(6)     & $4.5$    &  881    &  7058    &       & 10, 12, 14        \\
a09m220  &$48^3\times 96$  & 220.3(2)     & $4.71$   &  890    &  7120    &       & 10, 12, 14        \\
a09m130  &$64^3\times 96$  & 128.2(1)     & $3.66$   &  883    &  4824    & 56512 & 10, 12, 14        \\
\hline
a06m310  &$48^3\times 144$ & 319.3(5)     & $4.51$   &  865    &  3460    & 64000 & 16, 20, 22, 24    \\
a06m220  &$64^3\times 144$ & 229.2(4)     & $4.25$   &  650    &  1320    & 41600 & 16, 20, 22, 24    \\
\hline
\end{tabular}
\caption{Description of the nine ensembles at $a=0.12$, $0.09$,
  $0.06$~fm used in this study.  $N_\text{HP}$ denotes the number of
  measurements with high precision solves and $N_\text{AMA}$ with the
  AMA method.}
\label{tab:ensembles}
\end{table}

%


\vskip -30pt
\section{Excited-State Contamination}
\label{sec:ESC}

The goal is to extract all observables (charges, charge radii, form
factors, generalized parton distribution functions, TMDs) by
calculating matrix elements between ground-state nucleons.  Excited
state contamination is, however, a significant challenge to the
calculations of matrix elements within nucleon
states~\cite{Bhattacharya:2015wna}.  We employ three strategies to
control the ESC. (i) The overlap between the nucleon operator and the
excited states is reduced by using smeared sources when calculating
the quark propagators. (ii) We calculate the three-point correlation
functions for a number of values of the source-sink separation $t_{\rm
  sep}$ given in Table~\ref{tab:ensembles}. (iii) Data at various
$t_{\rm sep}$ are fit simultaneously using the two-state ansatz given
to estimate the $t_{\rm sep} \to \infty$ value as follows:
\begin{align}
C^\text{2pt} (t_f,t_i) = &  {|{\cal A}_0|}^2 e^{-M_0 (t_f-t_i)} + {|{\cal A}_1|}^2 e^{-M_1 (t_f-t_i)}\,, \nonumber \\
C^\text{3pt}_{\Gamma}(t_f,\tau,t_i)  = & 
    |{\cal A}_0|^2 \langle 0 | \mathcal{O}_\Gamma | 0 \rangle  e^{-M_0 (t_f-t_i)} +
   |{\cal A}_1|^2 \langle 1 | \mathcal{O}_\Gamma | 1 \rangle  e^{-M_1 (t_f-t_i)} \nonumber \\
   + & \ \ 
   {\cal A}_0{\cal A}_1^* \langle 0 | \mathcal{O}_\Gamma | 1 \rangle  e^{-M_0 (\tau-t_i)} e^{-M_1 (t_f-\tau)} + 
   {\cal A}_0^*{\cal A}_1 \langle 1 | \mathcal{O}_\Gamma | 0 \rangle  e^{-M_1 (\tau-t_i)} e^{-M_0 (t_f-\tau)} .
\label{eq:2pt_3pt}
\end{align}
The masses and amplitudes $M_0$, $M_1$, ${\cal A}_0$, and ${\cal A}_1$
of the ground and ``first'' excited states are obtained from fits to
the two-point functions.  These are then used as inputs in the fit to
the 3-point function to extract the three matrix elements $\langle 0 |
O_\Gamma | 0 \rangle$, $\langle 0 | O_\Gamma | 1\rangle $ and $\langle
1 | O_\Gamma | 1 \rangle$. Propagation of errors between the two fits is 
taken into account by doing both within the same jackknife process.

\begin{center}
\begin{figure*}
\vskip -0.2in
\hskip -0.4in
\includegraphics[width=1.15\textwidth]{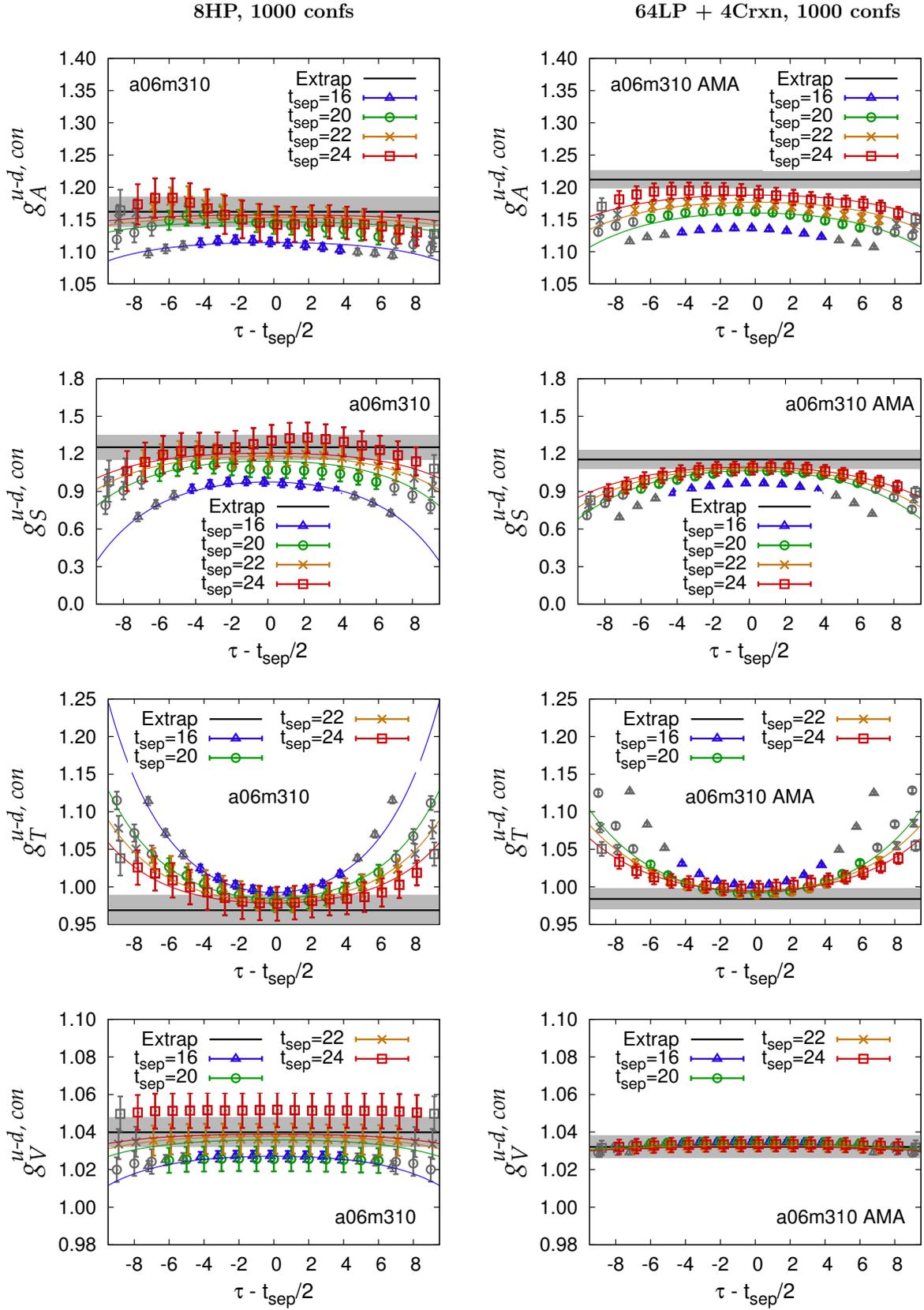} 
\vskip -0.2in
\caption{Figure illustrates reduction in errors between 8000 HP and 64,000 AMA measurements 
  for the same computation cost. We show data for $t_{\rm sep} = 16,20,22,24$ and the 
  $t_{\rm sep} \to \infty $ estimate given by the fit. }
\label{Fig:HPvsAMA}
\end{figure*}
\end{center}

Fig.~\ref{Fig:HPvsAMA} illustrates the improvement in the $a12m310$
ensemble data for the isovector charges on using the AMA method
compared to all HP.  The two estimates are consistent, with the errors
in the AMA data smaller by $\approx \sqrt8$, consistent with the
increase in the statistics: small enough to resolve the trend with
$t_{\rm sep}$. We find that the two state fit captures these trends,
significantly increasing the confidence in the fit to estimate the
$t_{\rm sep} \to \infty$ value.

Fits to 2- and 3-point functions using Eq.~\eqref{eq:2pt_3pt} yield
five physical quantities: the masses $M_0$ and $M_1$ and the three
matrix elements: the charge $\langle 0 | O_\Gamma | 0 \rangle$ and the
ME $\langle 0 | O_\Gamma | 1\rangle $ and $\langle 1 | O_\Gamma | 1
\rangle$, {\it albeit} ESC and discretization errors have to be
removed from each.  On the other hand, the two amplitudes, ${\cal
  A}_0$ and ${\cal A}_1$, depend on the nucleon interpolating operator
and the smearing used at the source and sink ends of the quark
propagator. From Eq.~\eqref{eq:2pt_3pt}, it is clear that to reduce
ESC one needs to reduce the ratio ${\cal A}_1/{\cal A}_0$. Our tests
show that increasing the size of the Gaussian smearing reduces ESC in
the charges, but beyond a certain size the errors in the correlation
functions start to increase. A good compromise choice for the smearing
parameter is $\sigma \approx 0.60$~fm, with $\sigma $ defined as
in~\cite{Bhattacharya:2015wna}. \looseness-1

Our conclusion is that ESC is as large as $15\%$ in our data, most
notable in $g_A$. ESC can be reduced significantly to $\sim 5\%$, by
choosing $\sigma \approx 0.60$~fm. $g_A$ and $g_S$ converge from
below, while $g_T$ from above, thus residual ESC would result in
underestimating $g_A$ and $g_S$ and overestimating $g_T$. With ESC under
control, a $\approx 3\%$ estimate of $t_{\rm sep} \to \infty$ value can be
obtained using the 2-state fit ansantz, Eq.~\eqref{eq:2pt_3pt}, with
data obtained at multiple values of $t_{\rm sep}$ over the range
$1-1.5$~fm.

\vskip -30pt
\section{Combined fits in lattice volume, spacing and quark mass}

\begin{figure}
\includegraphics[width=.99\textwidth]{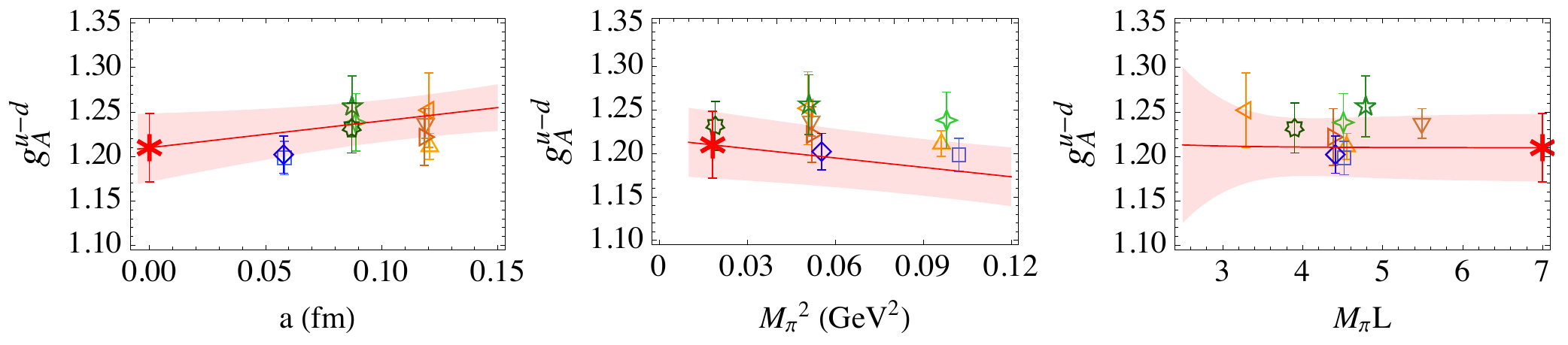} 
\includegraphics[width=.99\textwidth]{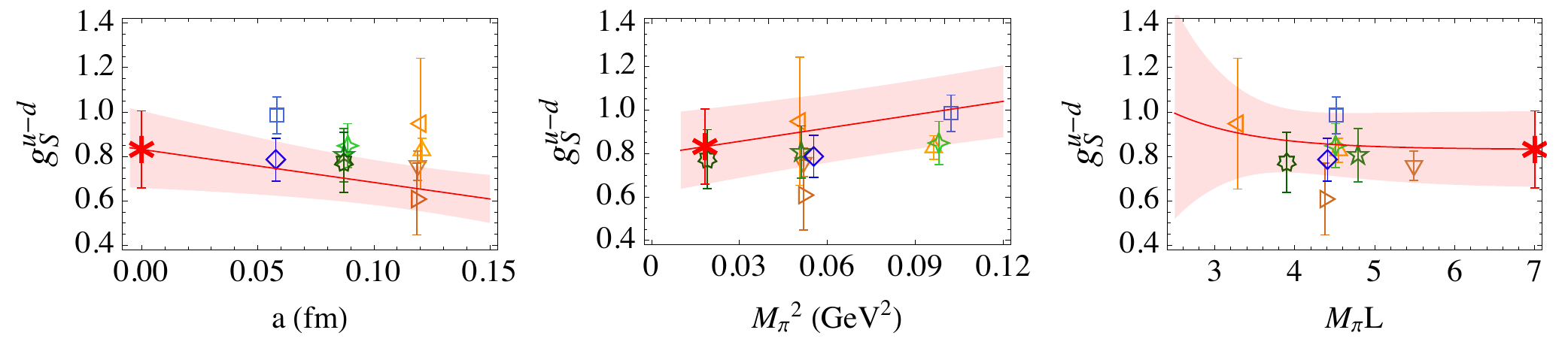} 
\includegraphics[width=.99\textwidth]{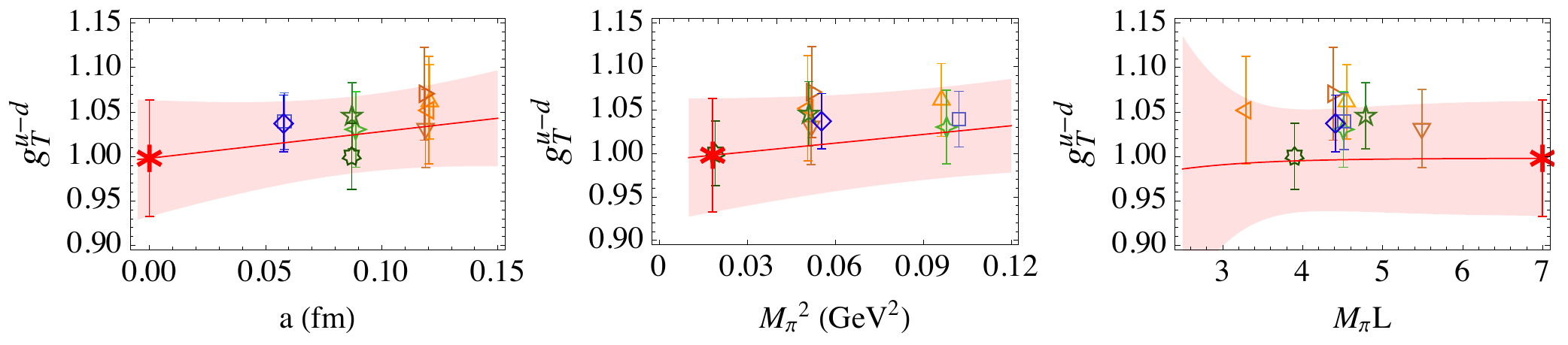} 
\vspace{-10pt}
\caption{Simultaneous fits in $a$, $M_\pi$ and $M_\pi L$ using
  Eq.~(\protect\ref{Eq:extrap}) to obtain the final results for the
  isovector charges $g_A^{u-d}$, $g_S^{u-d}$ and $g_T^{u-d}$.}
\label{Fig:extrap}
\end{figure}

The renormalization factor for the bilinear quark operators is
calculated using the RI-sMOM scheme~\cite{Martinelli:1994ty} and converted to the continuum
$\overline{MS}$ scheme at 2~GeV. With the renormalized charges
obtained at various values of $a$, $M_\pi$ and lattice volume $L$ in hand, the
final result in the $a \to 0$, $M_\pi \to 135$~MeV and $M_\pi L \to
\infty$ limits are obtained using the lowest order correction term in
each of the three variables (we are not able to explore higher order corrections with current data):
\begin{equation}
g(a, M_\pi, M_\pi L) = g^{{\rm physical}} + \alpha a + \beta M_\pi^2 + \gamma e^{-M_\pi L} \,.
\label{Eq:extrap}
\end{equation}
In Fig.~\ref{Fig:extrap}, we show the fit for the isovector charges
$g_A^{u-d}$, $g_S^{u-d}$ and $g_T^{u-d}$.  Our present best estimates
are based on a 7-point fit neglecting the $a12m220S$ and $a12m220$
ensemble data points and setting $\gamma=0$ ( $\gamma$ is ill
determined and no significant volume dependence is seen for $M_\pi L
\ge 4$) :
\begin{align}
g_A^{u-d} = & 1.21(3) \nonumber \\
g_T^{u-d} = & 1.005(59) \nonumber \\
g_S^{u-d} = & 0.95(12) \, .
\end{align}

{\it Prognosis for 2\% estimates}: Based on our current analyses, we
conclude that to extract $g_A$ and $g_T$ with $2\%$ uncertainty will
require $O(2000)$ configurations with $O(200,000)$ measurements on
each of the HISQ ensembles at $a = 0.09, 0.06$ and $0.045$~fm.
Estimates of $g_S$ with similar precision will require an order of
magnitude more measurements. Equivalently, Clover ensembles at
$a=0.09$ 0.07 and $0.05$~fm with similar values of $M_\pi \approx
300$, 200 and 140~MeV and $M_\pi L \ge 4$ would be needed. These are
being generated by the JLab/W\&M lattice group.

\vskip -10pt
\section{BSM contributions to Neutron Electric Dipole Moment}

\begin{figure}
\includegraphics[width=.45\textwidth]{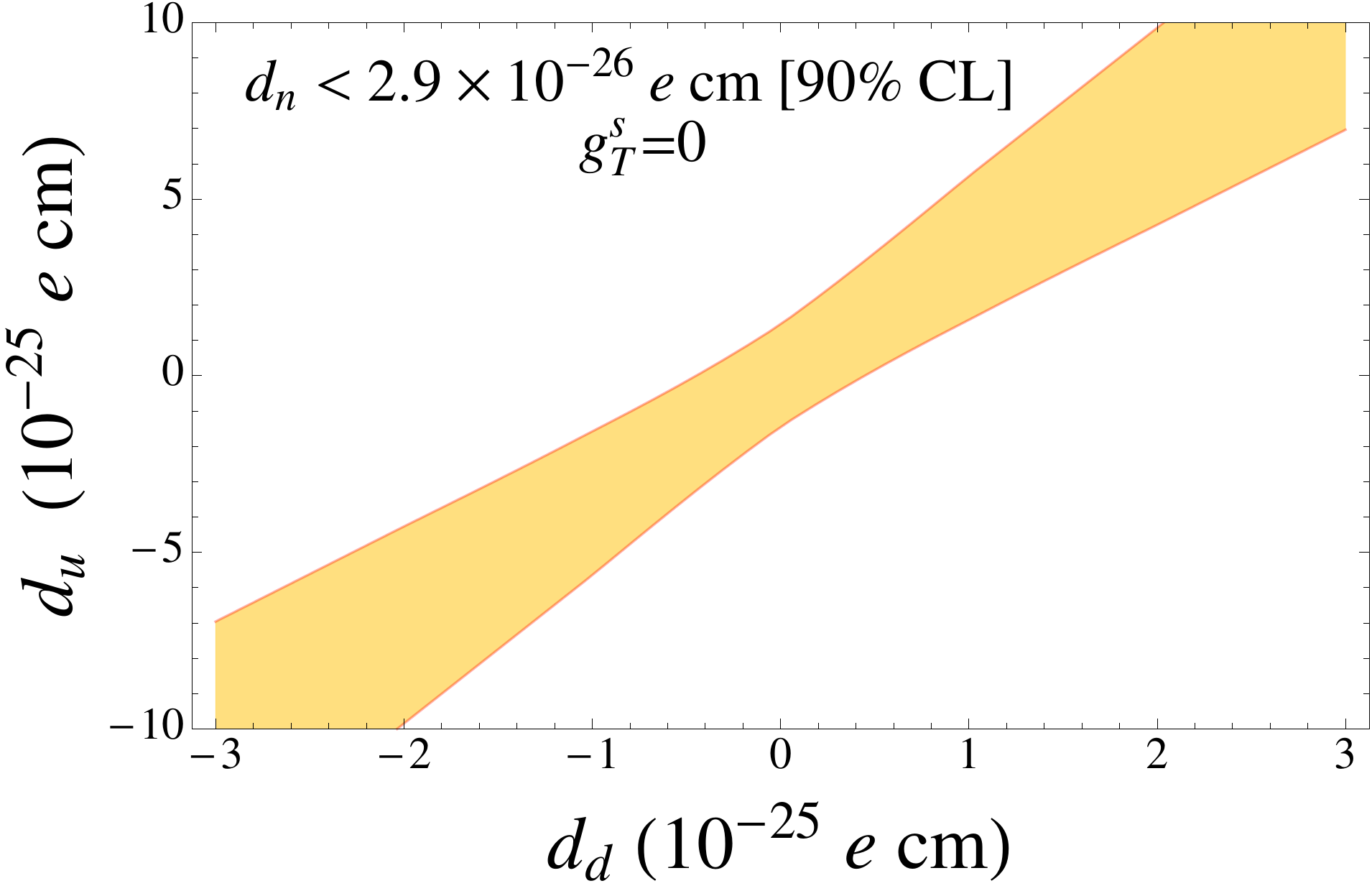} 
\hskip 20pt
\includegraphics[width=.45\textwidth]{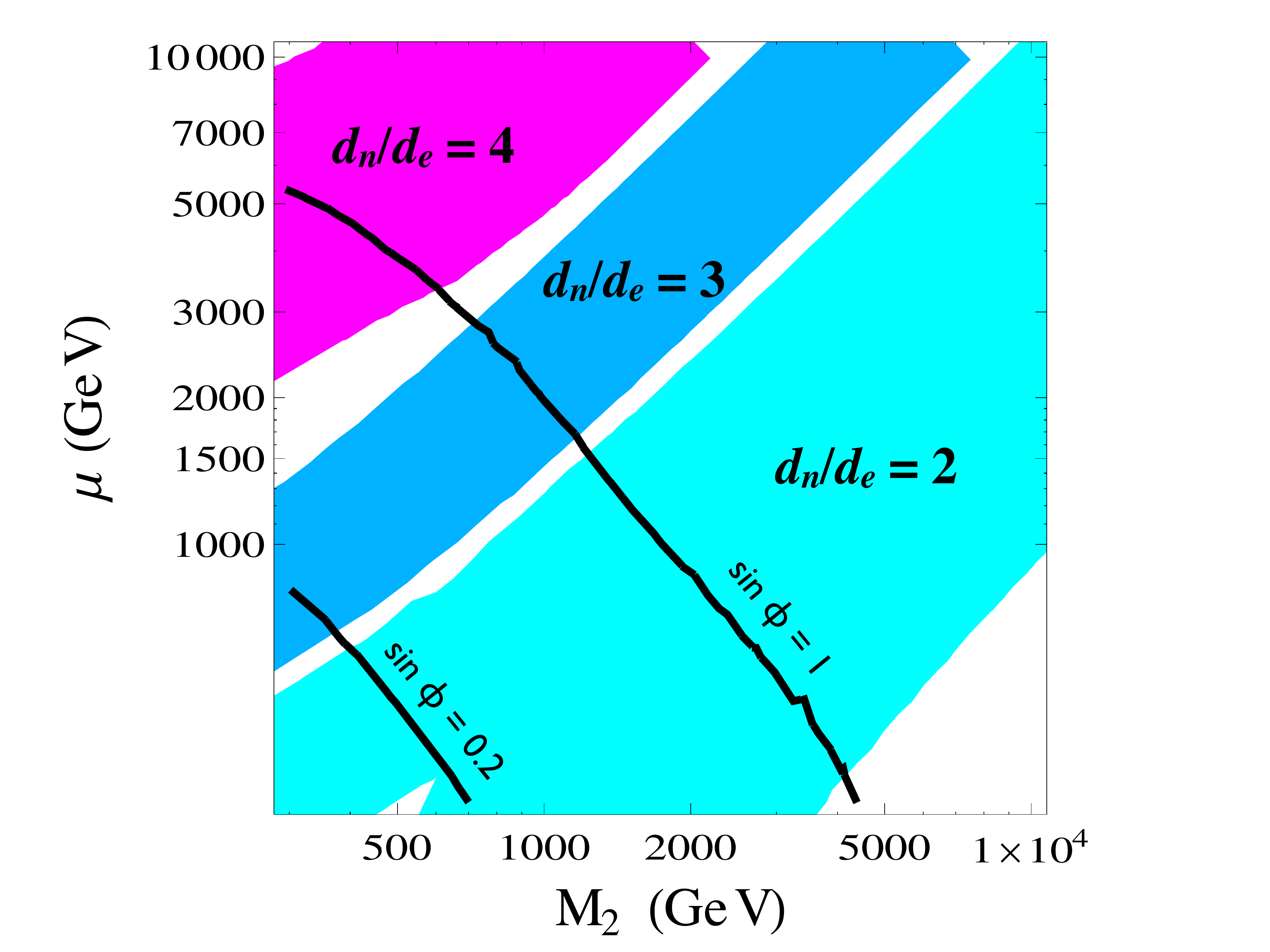} 
\vspace{-10pt}
\caption{(Left) Allowed region of $d_u$ and $d_d$ using the current bound on the neutron EDM 
and $g_T^q$ in Eq.~\protect\eqref{eq:gT}. 
(Right) Regions in $M_2$-$\mu$ plane corresponding to
    $d_n/d_e = 2, 3, 4$ in split SUSY, obtained by varying
    $g_T^{u,d,s}$ within our estimated uncertainties.  The lines
    correspond to $d_e = 8.7 \times 10^{-29} \, e$ cm for $\sin \phi 
    = 0.2, 1$.  }
\label{Fig:pheno}
\end{figure}

To explain weak-scale baryogenesis requires CP violation much larger
than present in the CKM matrix in the standard model. Most BSM
theories have additional sources of CP violation. In an effective
field theory formulation, the leading new operators that arise and contribute to the
EDM are the quark EDM $d_q \ \bar{q} \sigma_{\mu \nu} \gamma_5 F^{\mu \nu} q $ and
the quark chromo EDM $d_q \ \bar{q} \sigma_{\mu \nu} \gamma_5 {\tilde
  G}^{\mu \nu} q $~\cite{Pospelov:2005pr}.  The quark EDM contribution
of the light $u,d,s$ quarks to the neutron EDM $d_n$ is then given by
\begin{equation}
d_n = d_u g_T^u + d_d g_T^d + d_s g_T^s  \, .
\label{eq:nEDM}
\end{equation}
$d_q$ are the quark EDM induced in BSM theories and run down to
the low energy hadronic scale 2~GeV in the $\overline{MS}$ scheme.
The flavor diagonal tensor charges $g_T^q$, calculated using lattice
QCD, are also renormalized at the same point. The product $d_n$ is, therefore, 
scale and scheme independent.  The calculation of the connected part
of $g_T^q$ is the same as for isovector charges.  Estimates of the
disconnected contribution, obtained using the AMA method
in~\cite{Bhattacharya:2015wna}, have large errors and are consistent
with zero. For $g_T^u$ and $g_T^d$ we take the largest value on the
five ensembles simulated and add it as an additional error in the
connected contribution.  Estimates of $g_T^s$ are also consistent with
zero, however, in this case we were able to extrapolate to the
continuum limit. Using our results for the
neutron~\cite{Bhattacharya:2015wna} (note the quark label interchange
$u \leftrightarrow d$ between the neutron and proton) \looseness-1
\begin{align}
g_T^{u} = & -0.233(28) \, , \nonumber \\
g_T^{d} = & 0.774(66)  \, , \nonumber \\
g_T^{s} = & 0.008(9) \, ,
\label{eq:gT}
\end{align}
and the current bound $d_n < 2.9 \times 10^{-26} e$ cm, the bounds on
$d_u$ and $d_d$ are shown in Fig.~\ref{Fig:pheno} (left)~\cite{Bhattacharya:2015esa}.

In general, BSM theories generate a number of CP-violating
operators. However, in the ``split SUSY''
model~\cite{ArkaniHamed:2004fb}, fermion EDM operators provide the
dominant BSM source of CP violation. In this model, all scalars,
except for one Higgs doublet, are much heavier than the electroweak
scale.  This SUSY scenario achieves gauge coupling unification, has a
dark matter candidate, and avoids the most stringent constraints
associated with flavor and CP observables mediated by one-loop
diagrams involving scalar particles.
Contributions to fermion EDMs arise at two loops due to CP violating
phases in the gaugino-Higgsino sector, while all other operators are
highly suppressed~\cite{Giudice:2005rz}.  Using our estimates of
$g_T^q$, we updated the analysis following Ref.~\cite{Giudice:2005rz}.
For example, in Fig.~\ref{Fig:pheno} (right) we show allowed regions
in the gaugino ($M_2$) and Higgsino ($\mu$) mass parameter space for
different $d_n/d_e = 2, 3, 4$~\cite{Bhattacharya:2015esa}.  In this
model, the $d_q$ also depend on a phase $\phi$ and a factor ${\rm
  tan}\, \beta$. Using this analysis and the current 90\% C.L. limit on
the electron EDM $d_e = 8.7 \times 10^{-29} \, e$~cm, we derived an upper limit for the neutron EDM
in split SUSY i.e., $d_n < 4 \times 10^{-28} \, e$ cm. Consequently,
an observation of the neutron EDM between the current limit of $3
\times 10^{-26} \, e$ cm and $4 \times 10^{-28} \, e$ cm would falsify
the split-SUSY scenario with gaugino mass unification. \looseness -1


\vskip -40pt
\section*{Acknowledgments}
We thank the MILC Collaboration for sharing the 2+1+1 HISQ ensembles.
Simulations were performed using the Chroma software
suite~\cite{Edwards:2004sx} on LANL Institutional Computing platforms; 
the USQCD Collaboration computers at Fermilab funded by the U.S. DoE; 
and a DOE ERCAP allocation at NERSC. RG, TB and BY are supported by DOE grant
DE-KA-1401020 and the LDRD program at LANL. HL is supported by the 
M. Hildred Blewett Fellowship from the APS. The nEDM analysis is being
done in collaboration with E. Mereghetti.


\vskip -30pt

\begin{thebibliography}{99}

\bibitem{Bhattacharya:2011qm}
T. Bhattacharya, $et al.$, %
Phys.Rev. {\bf D85} (2012) 054512. 

\bibitem{Bhattacharya:2015wna}
T. Bhattacharya, $et al.$, %
Phys.Rev. {\bf D92} (2015) 094511. 

\bibitem{Bhattacharya:2012nEDM}
T. Bhattacharya, $et al.$, %
PoS (LATTICE 2012), 179 (2012) and PoS (LATTICE 2013) 299 (2103).

\bibitem{Bazavov:2012xda}
A. Bazavov, et al., MILC Collaboration, Phys. Rev. {\bf D87}  (2013) 054505. 

\bibitem{Blum:2012uh}
T. Blum, T. Izubuchi, and E. Shintani, Phys.Rev. {\bf D88} (2013)  094503


\bibitem{Bhattacharya:2013ehc}
T. Bhattacharya, $et al.$, %
Phys.Rev. {\bf D89} (2014) 094502. 

\bibitem{Martinelli:1994ty}
G. Martinelli, et al., %
Nucl.Phys. {\bf B445} (1995) 81; 
C. Sturm, et al., 
Phys.Rev. {\bf D80} (2009) 014501. 

\bibitem{Bhattacharya:2015esa}
T. Bhattacharya, et al., %
arXiv:1506.04196 [hep-lat]

\bibitem{Pospelov:2005pr}
M. Pospelov and A. Ritz, Annals Phys. 318, (2005) 119. arXiv:hep-ph/0504231.

\bibitem{ArkaniHamed:2004fb}
N. Arkani-Hamed and S. Dimopoulos, JHEP 0506, (2005) 73. arXiv:hep-th/0405159.

\bibitem{Giudice:2005rz}
G. Giudice and A. Romanino, Nucl.Phys. B699 (2004) 65. arXiv:hep-ph/0406088.


\bibitem{Edwards:2004sx}
R. Edwards, B. Joo, Chroma Software System for LQCD, Nucl. Phys. Proc. Suppl. {\bf 140} (2005) 832

\end{thebibliography}
\end{document}